\documentclass[twocolumn]{aastex62}

\newcommand\ucla{UCLA Galactic Center Group, Physics and Astronomy Department, University of California, Los Angeles, CA 90024}
\graphicspath{{./}{figures/}}

\submitjournal{ApJL}

\shorttitle{Unprecedented variability of Sgr A* in NIR}
\shortauthors{Do et al.}

\begin{document}

\title{Unprecedented Near-Infrared Brightness and Variability of Sgr A*}

\correspondingauthor{Tuan Do}
\email{tdo@astro.ucla.edu}

\author{Tuan Do}
\affil{\ucla}

\author{Gunther Witzel}
\affil{Max Planck Institute for Radio Astronomy, Auf dem H\"ugel 69, D-53121 Bonn (Endenich), Germany}

\author{Abhimat K. Gautam}
\affil{\ucla}

\author{Zhuo Chen}
\affil{\ucla}

\author{Andrea M. Ghez}
\affil{\ucla}

\author{Mark R. Morris}
\affil{\ucla}

\author{Eric E. Becklin}
\affil{\ucla}

\author{Anna Ciurlo}
\affil{\ucla}

\author{Matthew Hosek Jr.}
\affil{\ucla}

\author{Gregory D. Martinez}
\affil{\ucla}

\author{Keith Matthews}
\affil{Division of Physics, Mathematics, and Astronomy, California Institute of Technology,\\
MC 301-17, Pasadena, California 91125, USA}

\author{Shoko Sakai}
\affil{\ucla}

\author{Rainer Sch\"odel}
\affil{Instituto de Astrof\'isica de Andaluc\'ia, Consejo Superior de Investigaciones Cient\'ificas,\\
Glorieta de la Astronom\'ia S/N, 18008 Granada, Spain}



\begin{abstract}

The electromagnetic counterpart to the Galactic center supermassive black hole, Sgr A*, has been observed in the near-infrared for over 20 years and is known to be highly variable. We report new Keck Telescope observations showing that Sgr A* reached much brighter flux levels in 2019 than ever measured at near-infrared wavelengths. 
In the K$^\prime$ band, Sgr A* reached flux levels of $\sim6$ mJy, twice the level of the previously observed peak flux from $>13,000$ measurements over 130 nights with the VLT and Keck Telescopes. We also observe a factor of 75 change in flux over a 2-hour time span with no obvious color changes between 1.6 $\micron$ and 2.1 $\micron$. 
The distribution of flux variations observed this year is also significantly different than the historical distribution. 
Using the most comprehensive statistical model published, the probability of a single night exhibiting peak flux levels observed this year, given historical Keck observations, is less than $0.3\%$. The probability to observe the flux levels similar to all 4 nights of data in 2019 is less than $0.05\%$. 
This increase in brightness and variability may indicate a period of heightened activity from Sgr A* or a change in its accretion state. It may also indicate that the current model is not sufficient to model Sgr A* at high flux levels and should be updated. 
Potential physical origins of Sgr A*'s unprecedented brightness may be from changes in the accretion-flow as a result of the star S0-2's closest passage to the black hole in 2018 or from a delayed reaction to the approach of the dusty object G2 in 2014. Additional multi-wavelength observations will be necessary to both monitor Sgr A* for potential state changes and to constrain the physical processes responsible for its current variability. 

\end{abstract}



\section{Introduction} \label{sec:intro}

The Galactic center hosts the closest supermassive black hole to the Earth, offering us a unique opportunity to study in detail the physical processes that occur in its vicinity. The Galactic black hole, Sgr A*, has been monitored extensively across many wavelength regimes
\citep[e.g.,][]{1974ApJ...194..265B,2000A&A...362..113F,2001Natur.413...45B,2002ApJ...577L...9H,2006ApJ...650..189Y,2008ApJ...682..373M,2010A&A...512A...2S,2013ApJ...774...42N,2015ApJ...802...69B,2016A&A...587A..37R,2017ApJ...845...35C,2017MNRAS.468.2447P,2018A&A...618L..10G}
For recent reviews see \cite{2010RvMP...82.3121G} and \cite{2012RAA....12..995M}.
These observations have shown that the source luminosity is 9 orders of magnitude below the Eddington luminosity and is highly variable \citep{1998ApJ...492..554N,1999ApJ...517L.101Q,2000ApJ...545..842Q,2000ApJ...539..809Q,2009ApJ...698..676D,2012A&A...537A..52E,2012AJ....144....1Y,2012A&A...540A..41H,2016A&A...589A.116M,2016MNRAS.461..552D,2017MNRAS.468.2447P}.

Observations of Sgr A* in the near-infrared are an effective way to monitor the variability of the black hole. First detected with adaptive optics (AO) images in 2003 \citep{2003Natur.425..934G,2004ApJ...601L.159G,2005ApJ...620..744G}, 
recent re-analysis of speckle imaging data has enabled detections of Sgr A* back to 1998, establishing a time baseline across 2 decades \citep{chen2019}. The near-infrared flux variations have been characterized as a red noise process that is correlated in time \citep[e.g.,][]{2009ApJ...691.1021D}. There have been several proposed models of the distribution of flux values over time: a single power-law model, a log-normal model \citep{2012ApJS..203...18W}, a log-normal model with an additional tail at higher flux levels \citep{2011ApJ...728...37D}, and a log-log normal distribution \citep{2014ApJ...791...24M}.
Recently, based on a comprehensive analysis of over 13,000 observations from historical AO data (2003--2014) from the Keck Telescopes and VLT and space data at 4.5 $\micron$ from Spitzer (2014--2017), \citet{2018ApJ...863...15W} found that the variability of Sgr A* in the near-infrared can be consistently described as a red-noise process with a single log-normal distribution for the flux variations. \citet{chen2019} found that the model from \citet{2018ApJ...863...15W} is also consistent with speckle data from the Keck Telescopes from 1995--2005. 

While historical near-infrared observations can all be fit with a single model, Sgr A* has the potential to greatly change its luminosity and variability. 
For example, observations of X-ray light echos from 6.4 keV iron line emission  suggest that in the past few hundred years, Sgr A* may have undergone a few relatively brief (up to $\sim$10 yr) luminosity excursions by factors up to 10$^5$ \citep[e.g.,][]{1996PASJ...48..249K,2013A&A...558A..32C,2018A&A...612A.102T}. 
In addition, the environment around Sgr A* is very dynamic, with stars and other objects passing near the black hole, which may affect its accretion flow. In 2018, the star S0-2 reached within 100 AU of the black hole. Also, in recent decades, two dusty objects (G1 \& G2) have shown signs of tidal interaction with the black hole \citep[e.g.][]{2012Natur.481...51G,2013ApJ...773L..13P,2013A&A...551A..18E,2014arXiv1410.1884W,2015ApJ...798..111P,2017ApJ...847...80W,2017ApJ...840...50P,2019ApJ...871..126G}. There have been numerous suggestions that these sources may deposit gas or alter the accretion onto Sgr A*, changing its luminosity and accretion state \citep[e.g.][]{2004MNRAS.350..725L,2012ApJ...755..155S}.

Here, we report new Keck Telescope near-infrared observations of Sgr A* in 2019. These observations show unusually bright flux levels and variability, with peak fluxes exceeding twice the maximum historical flux measurements. In Section \ref{sec:observations} we present the observations. Section \ref{sec:results} presents the light curves and analyses. Section \ref{sec:discussion} presents comparisons with the historical data, comparisons with models, and discussion of potential physical explanations for these observations. We conclude in Section \ref{sec:conclusions}.

\section{Observations}
\label{sec:observations}
\begin{figure*}
\centering
\includegraphics[width=6.5in]{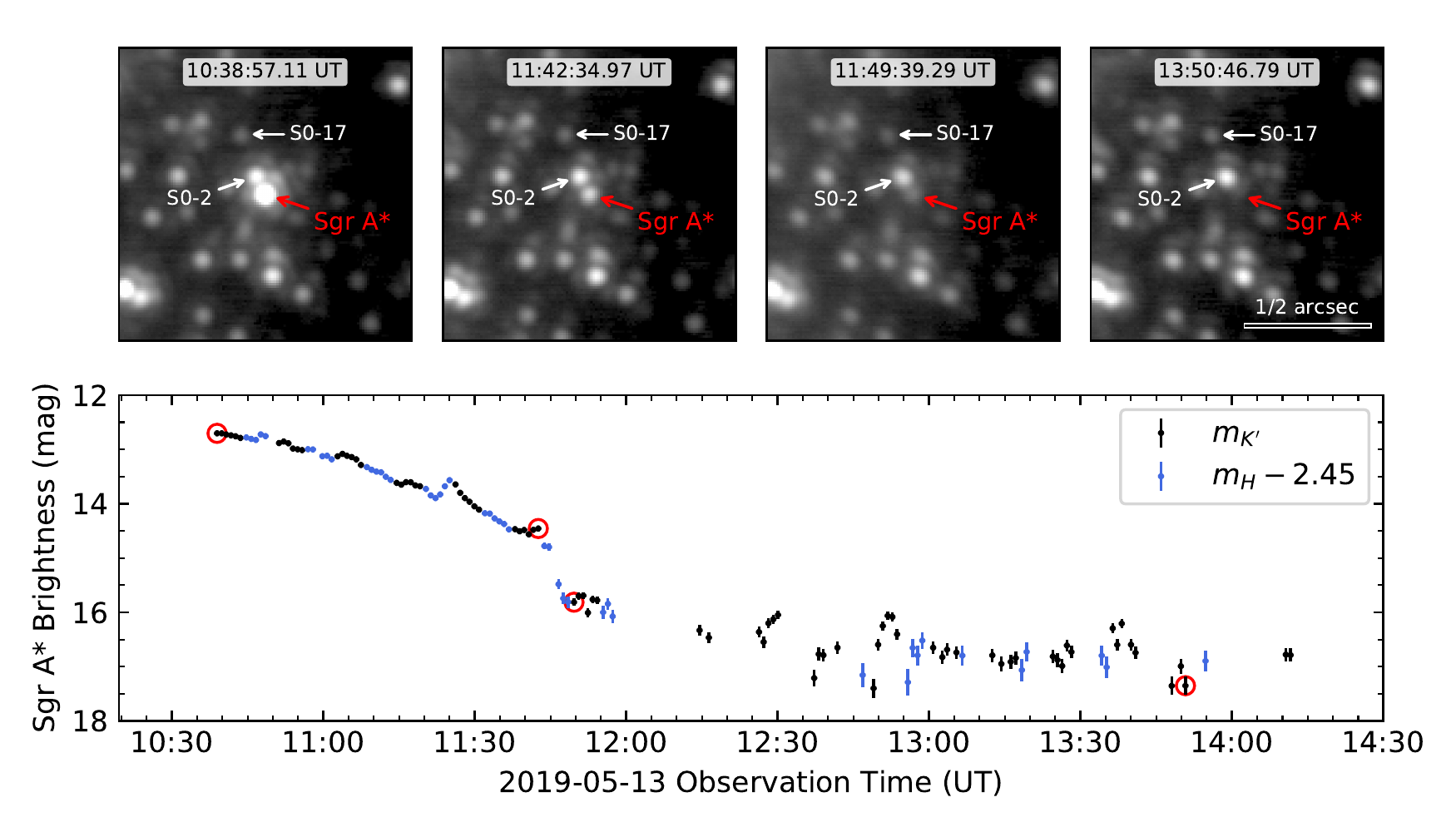}
\caption{\textbf{Top:} A series of K$^\prime$ images taken on 2019 May 13 centered on Sgr A* showing the large variations in brightness throughout the night. The first image is the brightest measurement ever made of Sgr A* in the near-infrared. Also labeled are nearby stars S0-2 (K$^\prime$ = 14 mag) and S0-17 (K$^\prime$=16 mag) for comparison. 
\textbf{Bottom:} K$^\prime$ (black) and H-band light curves of Sgr A* from 2019 May 13. On this night, we alternated between H and K$^\prime$ observations. The H-band magnitudes are offset using H$-$K$^\prime$ = 2.45 mag. There appear to be no significant color changes during the large change in brightness. Red circles show the location of the 4 images in the panels above.}
\label{fig:color}
\end{figure*}

We observed the Galactic center on 4 nights in 2019 with the Keck 2 Telescope using the narrow camera in the Near-Infrared Camera 2 (NIRC2) instrument with the Laser-guide Star Adaptive Optics (LGS AO) system \citep{2006PASP..118..297W}. 
We include all available 2019 observations from Keck having sampling duration $>20$ min and high enough image quality (full-width half-maximum $<100$ mas) to detect Sgr A*. 
We use the K$^\prime$ filter (2.12 $\micron$) on 3 nights and a combination of K$^\prime$ and H-band (1.64 \micron) filters on 1 night (Table \ref{tab:observations}). 
Individual K$^\prime$ images consist of 10 coadds of 2.8 s integration time each, while the H-band images consist of 4 coadds of 7.4 s each.
On 2019 May 13, the observations alternated between 6 frames of K$^\prime$ images and 6 frames of H-band images, for a total of about 6 minutes spent on each filter before switching. 
Images centered on Sgr A* are shown in Figure \ref{fig:color}. 
Standard image reduction methods were applied to the images including flat-fielding, sky subtraction, and cosmic-ray removal \citep[e.g.][]{2019ApJ...873....9J}. 

We extracted photometry for point sources in each individual image to construct light curves of Sgr A* and calibration stars. The brightness and position of all sources were measured using the PSF-fitting software StarFinder \citep{2000A&AS..147..335D}. 
Details on running StarFinder on individual frames for Galactic center observations can be found in \citet{2009ApJ...691.1021D}. Sgr A* is detected in almost every frame of observations on 3 nights. There are fewer detections of Sgr A* on 2019 May 19 due to lower-quality AO correction and faint flux levels for Sgr A*.  We calibrate the point source photometry using reference stars defined by \citet{2019ApJ...871..103G}. We convert between K$^\prime$ magnitudes to absolute fluxes F, using the relationship $F_{K^\prime} = 6.86\times10^5\times 10.0^{-0.4 K^\prime}$ mJy \citep{2005PASP..117..421T}. To convert from K$^\prime$ fluxes to K$_{s}$ fluxes, we use the filter transformation $F_{K_s} = 1.09 F_{K^\prime}$ computed for the color of Sgr A* \citep{chen2019}. We will mainly use observed fluxes in this work to avoid confusion with the value of extinction to apply. To de-redden the flux measurements, one can use the relationship $F_{deredden} = F_{obs} \times 10^{0.4 A_{K_s}}$. To compare to fluxes from \citep{2012ApJS..203...18W}, use $A_{K_s} = 2.46$ \citep{2010A&A...511A..18S}.

We estimate the Sgr A* relative photometric uncertainties by using the relationship between flux level and flux uncertainty of nearby stars. The photometric uncertainties for stars on each night were determined by using the standard deviation of their individual flux measurements. We find that the reference stars are stable during the nights with $<3\%$ photometric errors. Following \citet{2009ApJ...691.1021D}, we fit a power-law to the relationship between flux level and flux uncertainty for stars within 1$\arcsec$ of Sgr A*. We then use this relationship to infer the flux uncertainty for the all flux measurements of Sgr A*. The Sgr A* relative photometric uncertainties are typically less than 5\% at high flux levels and about 15\% for faint flux values. 

\begin{deluxetable*}{lrrrrrr}[th]
\tablecolumns{7}
\centering
\tablecaption{Near-Infrared Sgr A* Observations}
\tablehead{\colhead{Date (UT)} & \colhead{N$_{obs}$} & \colhead{Duration} & \colhead{Max F$_{obs}$\tablenotemark{a}} & \colhead{Min F$_{obs}$} & \colhead{Max F$_{deredden}$\tablenotemark{b}} & \colhead{Min. F$_{deredden}$} \\
\colhead{} & \colhead{} & \colhead{(min)} & \colhead{(mJy)} & \colhead{(mJy)} & \colhead{(mJy)} & \colhead{(mJy)} }
\startdata
2019-04-19 & 35 & 147 & $  0.48 \pm   0.04 $ & $   0.09 \pm   0.01 $ &    4.6   &    0.85  \\
2019-04-20 & 152 &  87 & $  1.74 \pm   0.04 $ & $   0.07 \pm   0.01 $ &   16.7   &    0.65  \\
2019-05-13\tablenotemark{c} & 82 & 149 & $  6.19 \pm   0.08 $ & $   0.08 \pm   0.01 $ &   59.6   &    0.79  \\
2019-05-23 & 109 & 213 & $  0.70 \pm   0.02 $ & $   0.05 \pm   0.01 $ &    6.7   &    0.48  
\enddata
\tablenotetext{a}{Observed fluxes are converted from K$^\prime$ to K$_s$ filter.}
\tablenotetext{b}{Fluxes are dereddened using an extinction of $A_{K_s} = 2.46$ mag.}
\tablenotetext{c}{H-band observations are also made this night (Fig. \ref{fig:color}).}
\label{tab:observations}
\end{deluxetable*}

\begin{figure}

\centering
\includegraphics[width=3.5in]{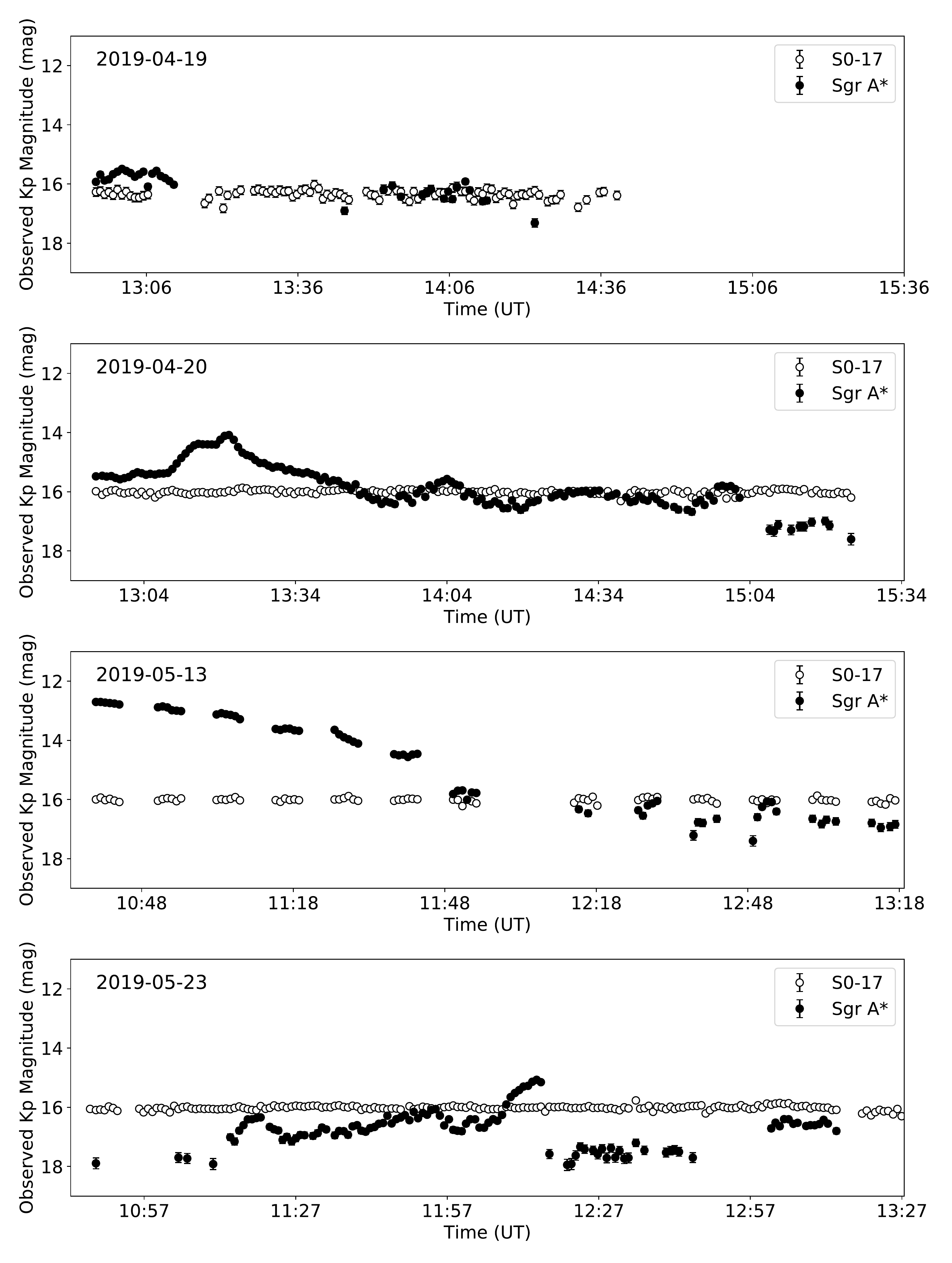}
\caption{K$^\prime$ light curves of Sgr A* (black) and a comparison star, S0-17 (white, located about 0.2$\arcsec$ from Sgr A*), on 4 nights of observations in 2019. We use stars within 1$\arcsec$ of Sgr A* to characterize the photometric error at different Sgr A* brightness levels. The photometric uncertainties are typically less than 5\% at high flux levels.}
\label{fig:lightcurves}
\end{figure}

\section{Results}
\label{sec:results}
\begin{figure*}
\centering
\includegraphics[width=6.25in]{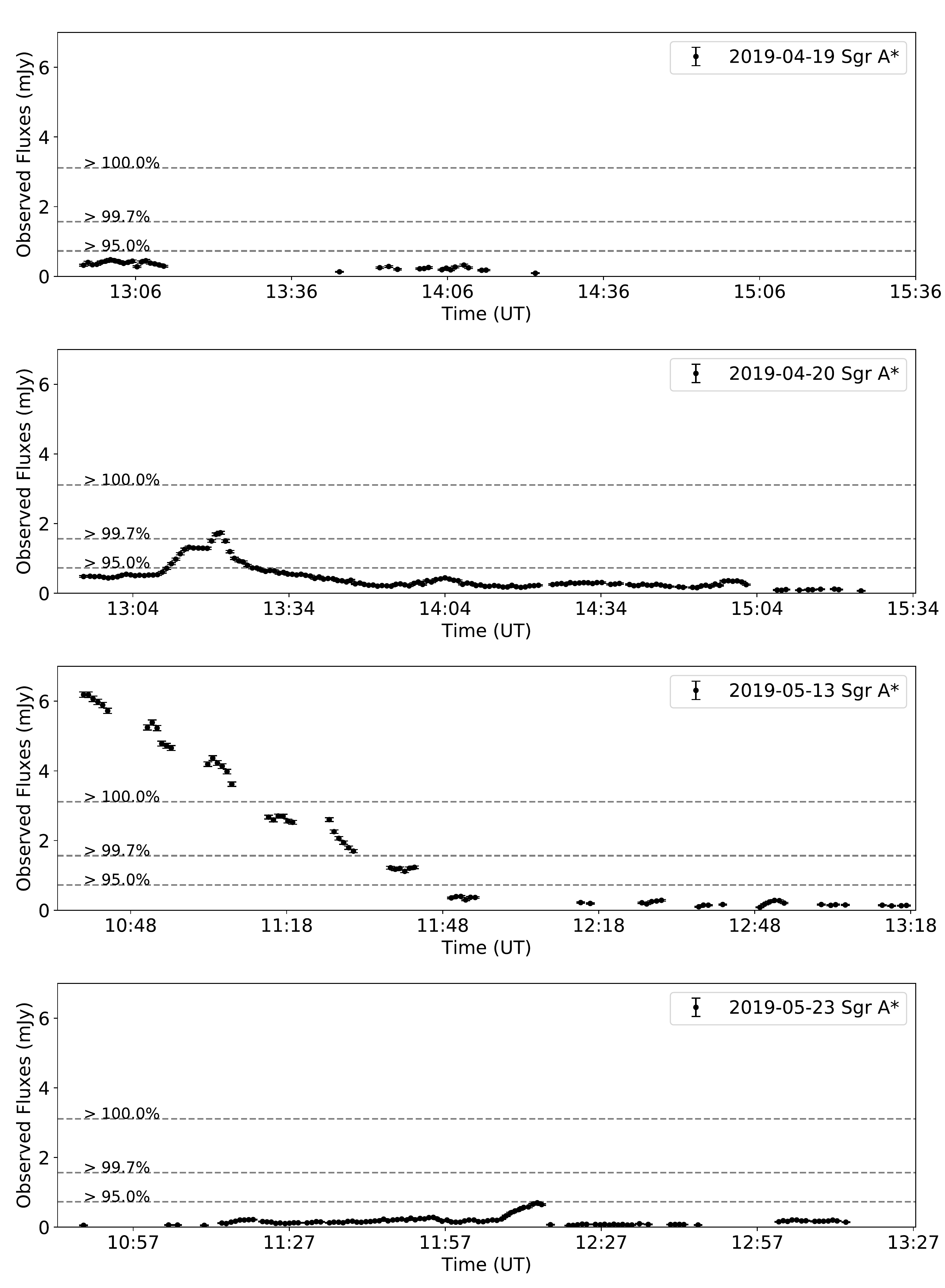}
\caption{Light curves of Sgr A* (black) obtained in 4 nights of observations in 2019 in observed flux units (in the K$_s$ filter). Dashed lines show the percentage of fluxes fainter than that level from historical data -- the 100\% line shows the maximum previously flux observed \citep{2011ApJ...728...37D,2018ApJ...863...15W}. 2019 May 13 shows flux levels exceeding the maximum historical data by a factor of 2, while 2019 April 20 show flux levels exceeding 99.7\% of previous observations. The light curve from 2019 May 13 falls linearly with time beginning with the first measurement. It likely that the peak flux level was even higher at earlier times.}
\label{fig:lightcurves_flux}
\end{figure*}

Our observations show Sgr A* to be highly variable in 2019. On 2019 May 13, Sgr A*'s flux level changed by a factor of 75 within 2 hours (from $6.19\pm 0.08$ mJy to $0.08\pm0.01$ mJy). 
The maximum observed fluxes occurred during the beginning of the observations, suggesting that Sgr A* was likely even brighter earlier in the night. On the night of 2019 April 20, Sgr A* flux also shows large variations with measurements ranging from $1.74\pm0.04$ mJy to $0.07\pm0.01$ mJy. The two other nights of our study  (2019 April 19 and 2019 May 23) show less variation at the level of 1 to 3 magnitudes of change during the night. Figures \ref{fig:color}, \ref{fig:lightcurves} and \ref{fig:lightcurves_flux} show the observed light curves.

Observations on 2019 May 13 also include H-band observations, which allows us to constrain the $H - K^\prime$ color of Sgr A* on this night. We find that the H-band and K$^\prime$ light curves can be matched using a color of $H - K^\prime$ = 2.45 mag. With this color shift, the H-band and K$^\prime$ points generally transition smoothly between the two filters. The exception may be during the large drop in flux at 11:42 UT. While out of the scope of this paper, we plan to characterize possible color changes during this night in a future work (Witzel et al. in prep). 

In addition to high flux values, two of the nights also show large drops in brightness over very short time scales. On 2019 May 13 at around 11:42 UT, the brightness of Sgr A* dropped from K$^\prime$ = 14.5 mag to K$^\prime$ = 15.8 mag, a factor of 3 in flux, within 7 min. 
An even larger change occurred on 2019 May 23 12:15 UT, when Sgr A*'s brightness changed from K$^\prime$ = 15.2 mag to K$^\prime$ = 17.6 mag, a factor of 9 in flux, within 2 min (corresponding to a light travel of $\sim$3 Schwarzschild radii for a $4\times10^6$ $M_\odot$ black hole). 


\section{Discussion}
\label{sec:discussion}

\begin{figure}[ht]
\centering
\includegraphics[width=3.5in]{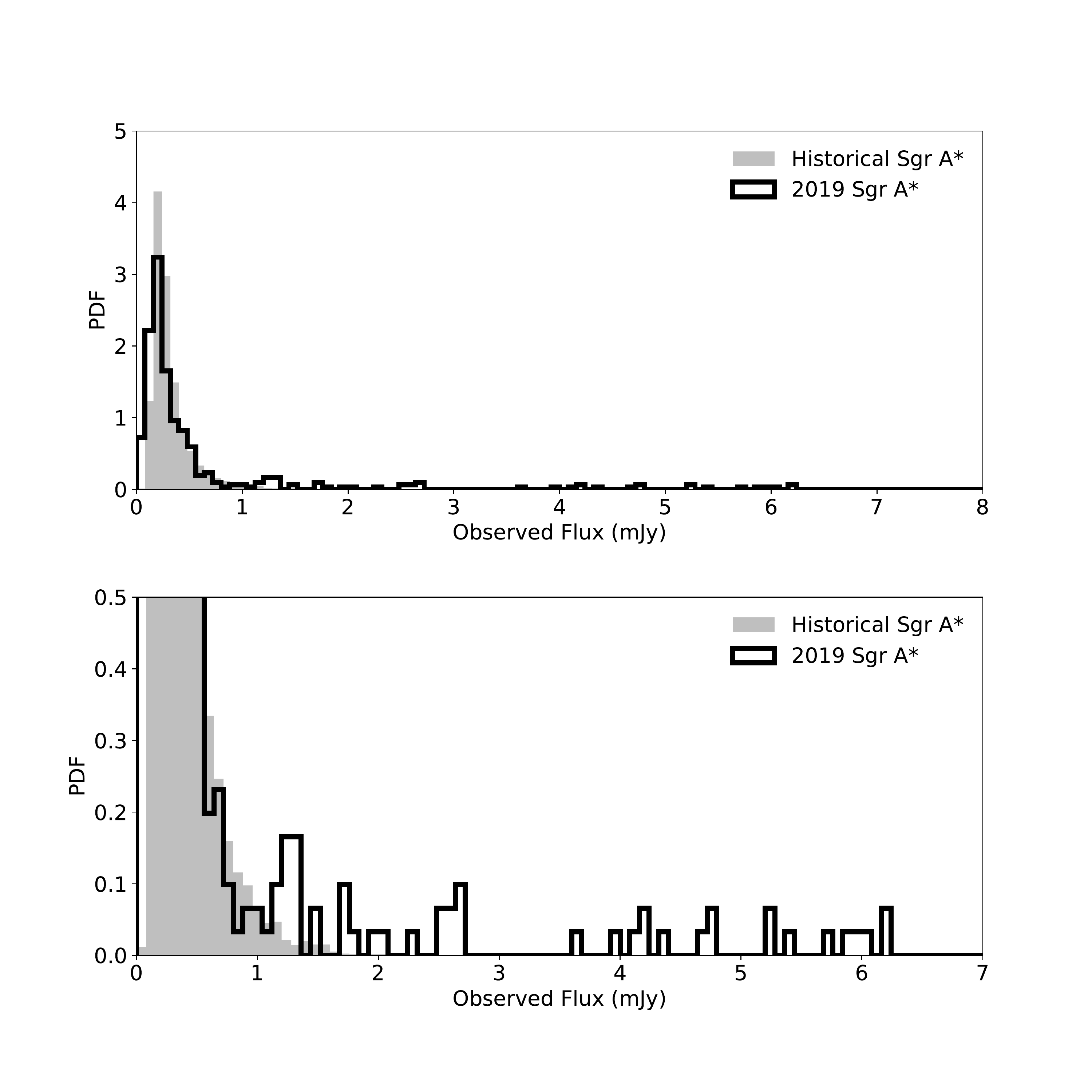}
\caption{Comparison of the distribution of Sgr A* flux variations from 2019 (black line) with the historical distribution (grey) from \citet{2018ApJ...863...15W}. Both distributions have been normalized to compare their shape and peaks. The bottom figure is a zoomed-in version of the top figure to show the tail of the distributions. A two-tailed KS-test shows that it is very unlikely for the two distributions to be drawn from the same underlying probability distribution.
}
\label{fig:distribution}
\end{figure}

The Sgr A*-IR observations presented here show peak flux levels that are unprecedented compared to the historical data. We use the distribution of flux variations from \citet{2018ApJ...863...15W}, which includes data spanning 10 years and over 13,000 $K_s$ flux measurements of Sgr A* from Keck and VLT (over 130 nights), to compare with the observations in 2019. The observations reported in \citet{2018ApJ...863...15W} were made with integration times between 28-40 s per measurement. The observations reported by \citet{chen2019} increase this time baseline to over 20 years. We find that the peak flux levels from 2019 May 13 exceed the maximum observed historical flux (3 mJy) by a factor of 2 (Figs. \ref{fig:lightcurves_flux} \& \ref{fig:distribution}). On 2019 April 20, the peak flux levels are brighter than 99.7\% of all historical data points. 

We also find the flux variations observed during the 4 observing periods in 2019 to be significantly different than in the historical data from \citet{2018ApJ...863...15W}. 
Using a two-tailed Kolmogorov-Smirnov (KS) test, we find a KS-statistic of 0.146, which corresponds to a probability of $\ll 0.01\%$ that a randomly drawn dataset based on the historical Sgr A*-IR distribution will produce a KS-statistic as extreme as, or more extreme than, the KS value derived when comparing the distribution of flux variations observed in 2019. 
The Anderson-Darling statistic \citep{scholz1987k}, which is more sensitive to the tails of the distribution has a value of 57. This also corresponds to a probability $\ll 0.01\%$ that a randomly drawn dataset based on the historical Sgr A*-IR distribution will produce such a statistic. 

\begin{figure}
\centering
\includegraphics[width=3.2in]{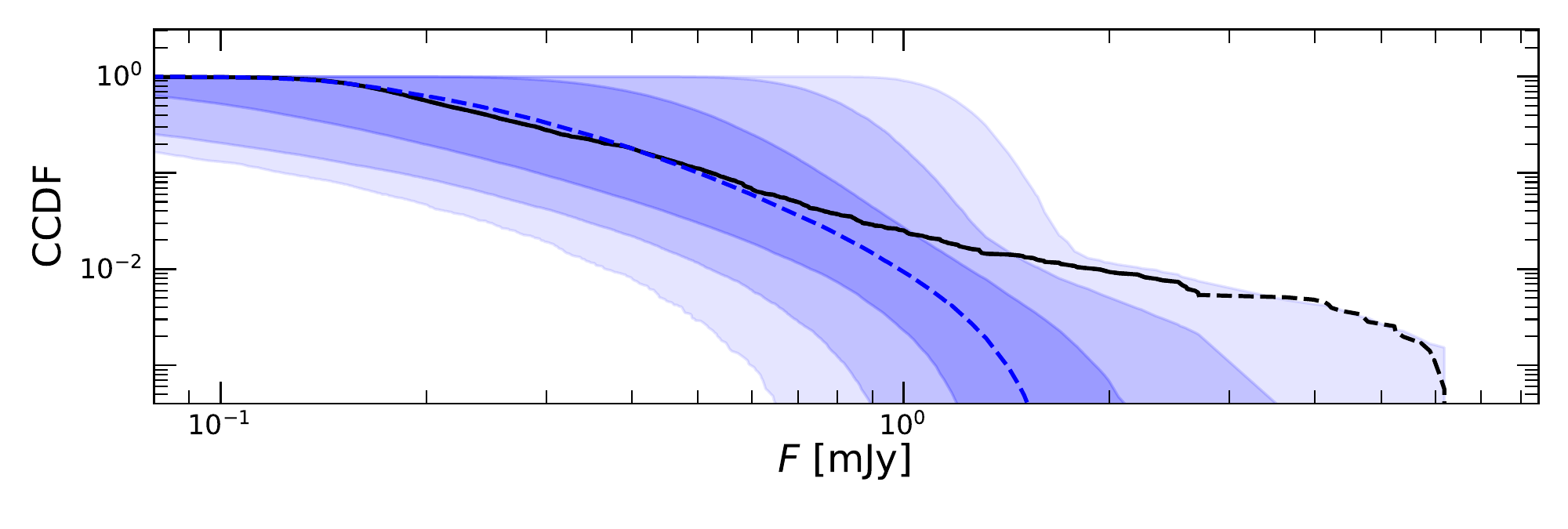}
\includegraphics[width=3.2in]{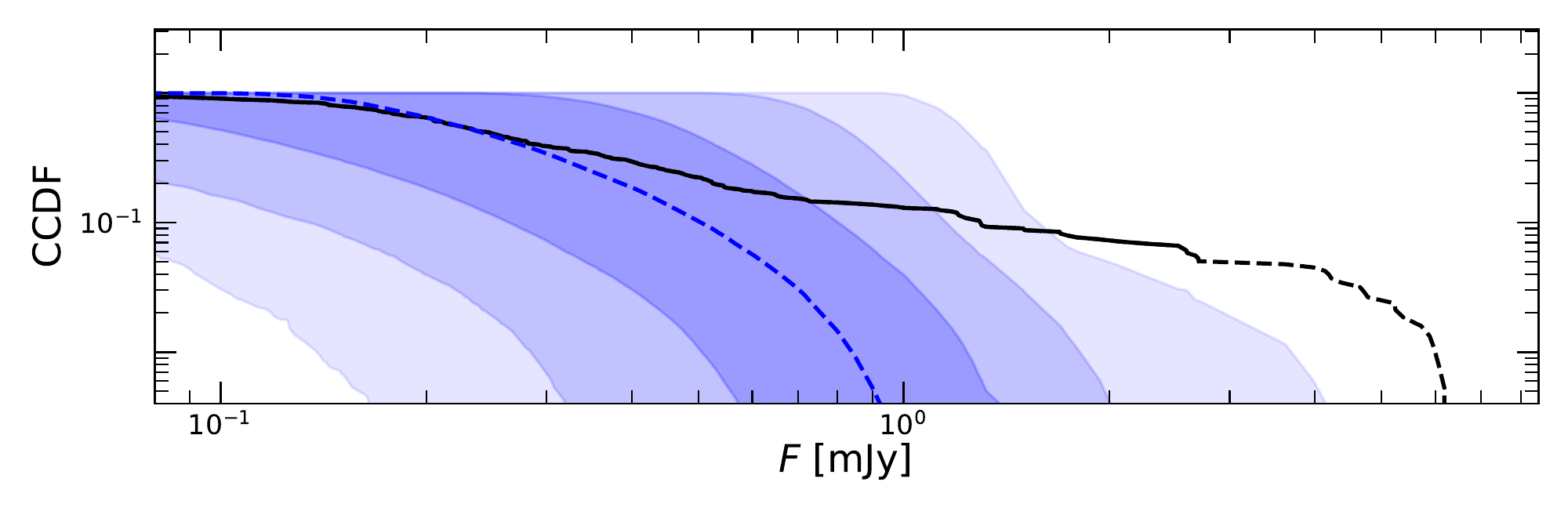}
\caption{\textbf{Top}: Comparison of the complementary cumulative distribution function (CCDF) of the observed data (historical data and 2019 data; black) and the median CCDF (dashed blue line) and the 1-, 2-, and 3-sigma contours calculated from 10,000 simulations. The simulations were drawn from the posterior in \citet{2018ApJ...863...15W}. The dashed section of the observed CCDF represents flux densities which occurred only during the brightest flux excursion on 2019 May 13. 
These simulations show that if we repeated the entire experiment with the time sampling of $30$ historical nights of Keck observations 10,000 times, then the probability of observing a single night with flux levels as high as 6 mJy is less than $0.3\%$. \textbf{Bottom}:
like upper panel but contours determined from simulations based only on the time sampling of the 4 nights in 2019. Because 3 of the 4 nights have elevated Sgr A* flux levels, if an experiment with 4 nights of observations were repeated 10,000 times, the probability of observing Sgr A* flux levels similar to the nights in 2019 would be less than $0.05\%$.}
\label{fig:CDFs}
\end{figure}

To better understand the variability of Sgr A* we use the statistical model of \citet{2018ApJ...863...15W} to assess the probability of observing light curves similar to the 4 presented here. We use the \citet{2018ApJ...863...15W} model for comparison because: (1) this model can accommodate the temporal correlation in the flux of Sgr A*, which is crucial for statistical analyses and Monte Carlo simulations \citep[e.g.,][]{2009ApJ...691.1021D}, and (2) this model was created using the largest sample of near-infrared data to date. The correlation of Sgr A*'s flux with time has an important consequences: the brighter Sgr A* is, the larger the changes in flux density will be. If Sgr A* is already at an elevated flux density level, the probability to observe an even brighter state is much larger than the time-averaged
probability for such a bright state. For continuous samples with a monitoring duration not significantly longer than the correlation timescale ($\sim 245$ min), this can result in large deviations of the sample distribution of flux densities from the underlying distribution. 
In order to investigate the consequences of the flux correlations in time, we use the simulation approach presented in \citet{2018ApJ...863...15W}. We use the posterior of their model 3, i.e., a log-normally distributed, red-noise process with a characteristic break timescale that successfully describes the historic VLT and Keck data in K-band and eight full days of Spitzer/IRAC data at $4.5\mu\rm{m}$. From this posterior distribution, we sample 10,000 parameter combinations and generate one random light curve with the time sampling of the observed Keck data ($\sim30$ nights of historical data and 4 nights of 2019 data) for each parameter set. We present the complementary cumulative distribution function (CCDF, or 1 - CDF) of these 10,000 light curves in Fig.~\ref{fig:CDFs} in the form of the median CDF and 1-,2-, and 3-sigma credible intervals. 

Based on simulations drawn from the Sgr A* flux model, we can compute the probability of observing the light curves presented here. 
The simulations show that if we repeated our experiment 10,000 times with the time sampling of the 30 nights of historical Keck observations (from 2005--2013) of Sgr A*, and including the 2019 nights, there is a $0.3\%$ probability that we observe a single night with flux levels as high as seen on 2019 May 13. 
These simulations are the most consistent with our current observations because they have very similar noise properties, observed duration, and timing. 
If we also include VLT nights for a total of over 130 nights, then the probability is less than $1.5$\% to observe flux values higher than 6.18 mJy on a single night. 
We can also consider the probability of randomly drawing 4 light curves similar to the observations in 2019. This probability is considerably lower at less than $0.05\%$.  
We also note that we have only observed the decay of the light curve on 2019 May 13, which suggests the actual maximum was likely even higher (Fig. \ref{fig:lightcurves_flux}).


Here we examine two possibilities for explaining the very unusual brightness and variation of Sgr A* observed this year: (1) the statistical models need to be changed or updated, and (2) there is a physical change in the accretion activity of Sgr A*. 
Based on the statistical model of \citet{2018ApJ...863...15W}, with the four nights of observations, there is a probability of less than $0.05\%$ to
observe flux levels $> 6$ mJy. 
The long tail of high flux levels, which occurred on multiple nights observed this year, is a strong indication that a log-normal distribution of flux variations may not be sufficient to describe the IR activity of Sgr A*. 
Before these 2019 observations, \citet{2018ApJ...863...15W} had shown that the NIR spectral properties at low flux densities can be explained by log-normally distributed flux densities in K- and M-band, though \citet{2011ApJ...728...37D} had suggested including an additional component to describe high flux levels. 
Before this year, the highest flux levels deviated from the median model expectation by only $<2 \sigma$ \citep{2018ApJ...863...15W}; the observations this year suggest the model should be re-derived to include the new data to determine if an additional component in the model is required.
Most models also assumes Sgr A* is stationary (with no time dependence in the model parameters). 
With additional measurements, it will become possible to robustly differentiate between changes in the physical state of Sgr A* and a stationary model \citep[e.g.][]{2011ApJ...728...37D,2014ApJ...791...24M}. We can also study potential changes in the activity of Sgr A* in the past several years by including data from Keck and VLT from 2014-2018 into an analysis similar to that of \citet{2018ApJ...863...15W}, which was based on data from 2003 to 2013. With additional data, we can also re-examine the timing characteristics and break-time scale for light curves observed this year to look for deviations from the current model (e.g. such as a shorter break-time scale) and to better understand the time evolution of accretion on Sgr A*. 

Another possibility is that these observations are an indication that Sgr A* is experiencing an increase in activity due to either a change in its accretion state or accretion rate. As mentioned above, the passage of windy stars through their periapses (most notably, S0-2, in May 2018) could potentially cause an extended rise in the accretion rate \citep{2004MNRAS.350..725L}, although \citet{2018MNRAS.478.3544R} argue that the effect of S0-2 on the structure of the accretion flow should be negligible. The S-star cluster has no known stars more massive than S0-2, so there are not any obvious high-mass-loss-rate candidate sources close to Sgr A*, including mass-losing giant stars.  An alternative mechanism for causing an accretion rate increase is simply the irregular gas flow toward SgrA*, which is likely to be quite lumpy and variable \citep{2008MNRAS.383..458C}.
Considerable work has recently been applied to this possibility by those interpreting the G sources as gas clumps orbiting toward Sgr A* \citep[e.g.,][]{2012ApJ...755..155S,2014ApJ...789L..33D,2015MNRAS.449....2M}.  Among them, \citet{2017PASJ...69...43K} predicted a radio and infrared brightening in the $\sim$2020 time frame.  








We predict that if the activity level of Sgr A* is indeed higher, then observations at other wavelengths should also show increased flux levels. \citet{2018ApJ...863...15W} can explain the log-normal nature of lower near-infrared flux densities by a radiative model that is dominated by an exponential synchrotron cooling cutoff within or near the near-infrared band. However, the authors note that this limits the flux densities to a range below 2 mJy. In order to reach higher flux densities (such as observed this year), the overall brightness of the synchrotron spectrum needs to scale accordingly. Once the emission is dominated by the synchrotron spectrum, there should be strong correlation in the fluxes at other wavelengths \citep[see also,][]{2012A&A...537A..52E}. 
This would predict large flux variations at radio, sub-millimeter, and X-ray wavelengths that would be observable and can be directly used to test this model.

\section{Conclusions}
\label{sec:conclusions}

Our recent observations of the Galactic center have captured Sgr A* in an unprecedented bright state in the near-infrared. Even more so, three of the four nights show Sgr A* in a clearly elevated state. The brightest flux levels observed in 2019 are over twice the peak flux value ever observed in the near-infrared from Keck and VLT. The distributions of flux variations from the four nights are also very unusual compared to the historical data, showing significant deviations from the model which was previously able to describe all historical Keck, VLT, \& Spitzer measurements \citep{2018ApJ...863...15W,chen2019}. 

The 2019 measurements push the limits of the current statistical models. These models may need to be revised to gain a better understanding of the probability of observing very high flux levels.
In addition, the statistical models for Sgr A* variability should be expanded to provide more robust tests for changes to the Sgr A* accretion properties over time. 

The major question is whether Sgr A* is showing increased levels of activity, and if so, how long it will last. Additional data, preferably multi-wavelength observations, throughout 2019 and beyond will be necessary to study the nature of its current variability. 

\acknowledgements
We thank the anonymous referee for their helpful comments. We thank the staff and astronomers at Keck Observatory especially Jim Lyke, Randy Campbell, Sherry Yeh, Greg Doppmann, Cynthia Wilburn, Terry Stickel, and Alan Hatakeyama. Support for this work was provided by NSF AAG grant AST-1412615, the W. M. Keck Foundation, the Heising-Simons Foundation, the Gordon and Betty Moore Foundation, the Levine-Leichtman Family Foundation, the Preston Family Graduate Fellowship (held by A.G.), and the UCLA Galactic Center Star Society. R. S. has received funding from the European Research Council under the European Union's Seventh Framework Programme (FP7/2007-2013) / ERC grant agreement number 614922. RS acknowledges financial support from the State Agency for Research of the Spanish MCIU through the ``Center of Excellence Severo Ochoa" award for the Instituto de Astrof\'isica de Andaluc\'ia (SEV-2017-0709). This research was based on data products from the Galactic Center Orbits Initiative, which is hosted at UCLA and is a key science program of the Galactic Center Collaboration. The W.M. Keck Observatory is operated as a scientific partnership among the California Institute of Technology, the University of California, and the National Aeronautics and Space Administration. The Observatory was made possible by the generous financial support of the W. M. Keck Foundation. The authors wish to recognize that the summit of Maunakea has always held a very significant cultural role for the indigenous Hawaiian community. We are most fortunate to have the opportunity to observe from this mountain.  
\bibliography{sgra.bib}

\end{document}